\documentclass[trackchanges,twocolumn]{aastex7}
\usepackage{graphicx}
\usepackage{amsmath}

\begin{document}

\title{Dense Molecular Ring-like structure in gaseous CO depletion region G34.74-0.12}
\shorttitle{Ring-like structure in G34.74-0.12}
\shortauthors{Lin et al. (2025)}

\author[0000-0001-5461-1905]{Shuting~Lin}
\affiliation{Department of Astronomy, Xiamen University, Zengcuo'an West Road, Xiamen, 361005}
\email{linst@stu.xmu.edu.cn}

\correspondingauthor{Siyi~Feng}
\email{syfeng@xmu.edu.cn}

\author[0000-0002-4707-8409]{Siyi~Feng}
\affiliation{Department of Astronomy, Xiamen University, Zengcuo'an West Road, Xiamen, 361005}
\email{syfeng@xmu.edu.cn}

\author[0000-0001-5950-1932]{Fengwei~Xu}
\affiliation{Kavli Institute for Astronomy and Astrophysics, Peking University, 5 Yiheyuan Road, Haidian District, Beijing 100871}
\affiliation{Department of Astronomy, School of Physics, Peking University, Beijing 100871}
\email{fengwei.astro@pku.edu.cn}

\author[0000-0002-7237-3856]{Ke~Wang}
\affiliation{Kavli Institute for Astronomy and Astrophysics, Peking University, 5 Yiheyuan Road, Haidian District, Beijing 100871}
\affiliation{Department of Astronomy, School of Physics, Peking University, Beijing
100871}
\email{kwang.astro@gmail.com}

\author[0000-0002-7125-7685]{Patricio~Sanhueza}
\affiliation{Department of Astronomy, School of Science, The University of Tokyo, 7-3-1 Hongo, Bunkyo, Tokyo 113-0033, Japan}
\email{patosanhueza@gmail.com}

\author[0000-0001-6106-1171]{Junzhi~Wang}
\affiliation{School of Physical Science and Technology, Guangxi University, Nanning 530004}
\email{junzhiwang@gxu.edu.cn}

\author[0000-0002-7299-2876]{Zhi-Yu~Zhang}
\affiliation{School of Astronomy and Space Science, Nanjing University, Nanjing 210093}
\affiliation{Key Laboratory of Modern Astronomy and Astrophysics (Nanjing University), Ministry of Education, Nanjing 210093}
\email{zzhang@nju.edu.cn}

\author[0000-0001-7511-0034]{Yichen~Zhang}
\affiliation{Department of Astronomy, Shanghai Jiao Tong University, 800 Dongchuan Rd., Minhang, Shanghai 200240}
\email{yichen.zhang@sjtu.edu.cn}

\author[0000-0002-6752-6061]{Kaho~Morii}
\affiliation{Center for Astrophysics $|$ Harvard \& Smithsonian, 60 Garden Street, Cambridge, MA 02138, USA}
\email{kaho.morii@grad.nao.ac.jp}

\author[0000-0003-2300-2626]{Hauyu~Baobab~Liu}
\affiliation{Department of Physics, National Sun Yat-Sen University, No. 70, Lien-Hai Road, Kaohsiung City 80424}
\affiliation{Center of Astronomy and Gravitation, National Taiwan Normal University, Taipei 116}
\email{hyliu.nsysu@mail.nsysu.edu.tw}

\author[0000-0003-4603-7119]{Sheng-Yuan~Liu}
\affiliation{Institute of Astronomy and Astrophysics, Academia Sinica, 11F of Astronomy-Mathematics Building, AS/NTU No. 1, Section 4, Roosevelt Road}
\email{syliu@asiaa.sinica.edu.tw}

\author[0000-0002-6540-7042]{Lile~Wang}
\affiliation{Kavli Institute for Astronomy and Astrophysics, Peking University, 5 Yiheyuan Road, Haidian District, Beijing 100871}
\affiliation{Department of Astronomy, School of Physics, Peking University, Beijing 100871}
\email{lilew@pku.edu.cn}

\author[0000-0002-6428-9806]{Giovanni Sabatini}
\affiliation{INAF, Osservatorio Astrofisico di Arcetri, Largo E. Fermi 5, I-50125, Firenze, Italy}
\email{giovanni.sabatini@inaf.it}

\author[0000-0002-1253-2763]{Hui~Li}
\affiliation{Department of Astronomy, Tsinghua University, Haidian DS 100084, Beijing}
\email{hliastro@tsinghua.edu.cn}

\author[0000-0003-3389-6838]{Willem Baan}
\affiliation{Xinjiang Astronomical Observatory, CAS, 150 Science 1-Street, Urumqi, Xinjiang 830011}
\affiliation{Netherlands Institute for Radio Astronomy ASTRON, NL-7991 PD Dwingeloo, The Netherlands}
\email{baan@astron.nl}

\author[0009-0001-7960-7512]{Zhi-Kai~Zhu}
\affiliation{School of Physical Science and Technology, Guangxi University, Nanning 530004}
\affiliation{Purple Mountain Observatory, Chinese Academy of Sciences, 10 Yuanhua Road, Nanjing 210023}
\email{zhuzhikai@tju.edu.cn}

\author[0000-0003-1275-5251]{Shanghuo Li}
\affiliation{School of Astronomy and Space Science, Nanjing University, Nanjing 210093}
\affiliation{Key Laboratory of Modern Astronomy and Astrophysics (Nanjing University), Ministry of Education, Nanjing 210093}
\email{shanghuo.li@gmail.com}


\begin{abstract}
We report the discovery of a dense molecular ring-like structure in a dense (10$^5$ cm$^{-3}$), cold (pc-scale CO depletion at a factor of 5), and young (10$^4$ year) star-forming region G34.74-0.12, revealed by C$^{18}$O~(2–1), HNC~(1–0), and N$_2$H$^+$~(1–0)
observations with the Atacama Large Millimeter/submillimeter Array (ALMA).  
The ring-like structure is redshifted with respect to the clump, spanning from $V_{\rm sys,lsr} + 0.9$ to $V_{\rm sys,lsr} + 2.9$ km~s$^{-1}$, with a total mass of 109~$M_{\odot}$. It is spatially coincident with 1.3~mm and 3.0~mm dust continuum emission from cores, and several protostellar outflows. However, no free-free emission or H\textsc{ii} region is detected in association with this structure.
With a slow expansion speed indicated by the position–velocity diagram, this ring structure differs from rings previously identified in more evolved star-forming regions.
Possible explanations for the ring-like structure include a relic wind-blown bubble produced by a deeply embedded young stellar object, a hollow cavity formed by cloud–cloud interactions, a gas ring resulting from a temperature gradient, or a line-of-sight superposition of multiple outflows or dense clouds.
This discovery offers a rare observational glimpse into the earliest dynamical processes involved in massive star formation. 

\end{abstract}

\keywords{\uat{Infrared dark clouds}{787} --- \uat{Star forming regions}{1565} --- \uat{Star formation}{1569} --- \uat{Interstellar medium}{847} --- \uat{Interstellar line emission}{844}}


\section{Introduction} 
\label{sec1:intro}

\setcounter{footnote}{0} 

Ring-like structures are commonly detected in star-forming regions, but their formation mechanisms appear to differ from case to case. Majority of them are linked to the activity of young stellar objects (YSOs) (e.g., \citealt{Arce2011ApJ...742..105A,Feddersen2018ApJ...862..121F,Duan2023ApJ...943..182D}) and the impact from nearby H\,\textsc{ii} regions or supernova feedback (e.g., \citealt{Churchwell2006ApJ...649..759C, Beaumont2014ApJS..214....3B, Watkins2023ApJ...944L..24W,Olguin2023ApJ...959L..31O,Shen2025A&A...693A..21S,Dewangan2025AJ....169...80D}).
Episodic ejections from YSOs can produce nearly spherical or arc-like structures, as traced by water maser emission \citep[e.g.,][]{Torrelles2001Natur.411..277T,Kim2013ApJ...767...86K}.
Projection effects of outflows viewed nearly along the pole-on axis and explosive outflows can also lead to apparent ring-like morphologies in molecular line emission (e.g., \citealt{Zapata2020ApJ...902L..47Z,Fernandez2020AJ....159..171F,Fernandez2021ApJ...913...29F}).
Other possible origins include large-scale streamers driven by gravitational accretion or turbulent flows (e.g., \citealt{Valdivia-Mena2022A&A...667A..12V,Mercimek2023MNRAS.522.2384M}), magnetically driven structures such as the so-called “magnetic wall” at the outer regions of a disk (e.g., \citealt{Machida2020MNRAS.494..827M,Tokuda2023ApJ...956L..16T}),   
as well as cloud–cloud collisions (CCC), which can trigger compression and accumulation of dense gas at the interface, forming cavities, arcs, or rings as a result of shock interaction \citep{Torii2015ApJ...806....7T,Fukui2018ApJ...859..166F}.
These ring-like structures are more difficult to identify than the relatively distinct features of outflows, primarily due to their complex morphologies \citep{Arce2011ApJ...742..105A}.

Most of the aforementioned ring structures have been identified in relatively evolved star-forming regions. In contrast, such structures are rarely reported in early-stage environments, particularly within infrared dark clouds (IRDCs), which are cold ($<$ 25~K), dense ($n$ $>$ 10$^5$ cm$^{-3}$), and serve as the cradles of the earliest stages of star formation \citep{Sanhueza2012ApJ...756...60S,Sanhueza2013ApJ...773..123S,Sanhueza2017ApJ...841...97S,Contreras2018ApJ...861...14C,Feng2019ApJ...883..202F,Li2019ApJ...878...29L,
Sanhueza2019ApJ...886..102S, Moser2020ApJ...897..136M,Morii2024ApJ...966..171M,Morii2025ApJ...979..233M}. 
Whether stellar feedback in the very early stages of star formation is sufficient to produce a ring structure on parsec scales remains an open question.
Detecting a pc-scale ring structure within a 70~$\mu$m dark region is therefore particularly important, as such a feature may offer valuable insights into the dynamical processes and physical conditions that shape the initial phases of star clusters.

We discover a dense molecular gas ring with parsec-scale extent in the 70-$\mu$m dark region G34.74$-$0.12. 
G34.74-0.12 is a filamentary high-mass star-forming region extending from northwest to southeast, located at a kinematic distance of 5.1$\pm$0.5~kpc \citep{2025arXiv250714564L}.
The systemic velocity with respect to the local standard of rest ($V_{\rm sys,lsr}$) is 79.0~km~s$^{-1}$, based on H$^{13}$CO$^+$(1-0) emission peak toward this region \citep[][Fig. A1]{Feng2020ApJ...901..145F}. 

In this letter, we characterize a dense molecular ring-like structure located in the gaseous CO depletion region G34.74$-$0.12, revealed at a linear resolution of $\sim$5000~au.
Section~\ref{sec:obser} presents the observations and data reduction. Section~\ref{sec:result} presents the observational results. Section~\ref{Sec:Analysis_Discussion} analyzes the physical and kinematic properties and discusses the possible origins of the ring-like structure. 
Section~\ref{sec:summary} summarizes our main conclusions.

\section{Observation AND DATA REDUCTION}
\label{sec:obser}
G34.74-0.12 has been observed as part of the ALMA Survey of 70 $\mu$m dark High-mass clumps in Early Stages (ASHES; Project ID: 2015.1.01539.S, 2017.1.00716.S, and 2018.1.00192.S, \citealt{Sanhueza2019ApJ...886..102S,Morii2023ApJ...950..148M}) at 1.3~mm, and Very Low Luminosity/Mass Ratio Clumps (VELLA) ALMA survey (Project ID: 2022.1.01203.S, Wang et al., in prep.) at 3.0~mm.

The ASHES observations used the 12~m array, the Atacama Compact 7~m Array (ACA), and the total power (TP) antennas.
For this region, ten 12~m array mosaics and three 7~m array mosaics were set to cover a 1$^{\prime}$ $\times$ 1$^{\prime}$ region centered at 18$^{\rm h}$55$^{\rm m}$09$^{\rm s}$.83, +01$^{\circ}$33$^{\prime}$14$^{\prime \prime}$.5 (ICRS). 
Two lines, C$^{18}$O~(2–1) (219.560 GHz) and H$_2$CO (3$_{0,3}$-2$_{0,2}$) (218.222~GHz) were targeted at a channel width of 488 kHz, corresponding to a spectral resolution of 1.33~km~s$^{-1}$ at the rest frequency of 219.560 GHz. 
Data calibration was carried out using the CASA versions 5.1.1 and 6.4.1. 
We use MIRIAD-based package, \emph{almica}\footnote{\href{https://github.com/baobabyoo/almica}{https://github.com/baobabyoo/almica}} \citep{Liu2015ApJ...804...37L}, to combine the 12m, ACA, and TP in the UV domain.
The synthesized beam of the C$^{18}$O~(2–1) line is 1$\farcs$68~$\times$~1$\farcs$24, with a position angle (P.A.) of -71.1$^\circ$.

The VELLA observation used two 12~m array configurations (C4 and C1). 
N$_2$H$^+$~($J$=1-0) was targeted at a channel width of 16~kHz, corresponding to a spectral resolution of 0.1~km~s$^{-1}$ at the rest frequency of 93.174~GHz. 
HNC~($J$=1-0) was targeted at a channel width of 31~kHz, corresponding to a spectral resolution of 0.21~km~s$^{-1}$ at the rest frequency of 90.664~GHz.
We combine data from two configurations and jointly image to achieve a maximum recoverable scale of $\sim30^{\prime\prime}$. The data calibration and imaging were performed using CASA 6.5.4. 
The synthesized beams of the N$_2$H$^+$~($J$=1–0) and HNC~($J$=1–0) lines are 1$\farcs$40 $\times$ 1$\farcs$30 (P.A. = 59.0$^\circ$) and 1$\farcs$44 $\times$ 1$\farcs$34 (P.A. = 66.5$^\circ$), respectively.

\section{Observational Results}
\label{sec:result}

\begin{figure*}
    \centering
    \includegraphics[width=\linewidth]{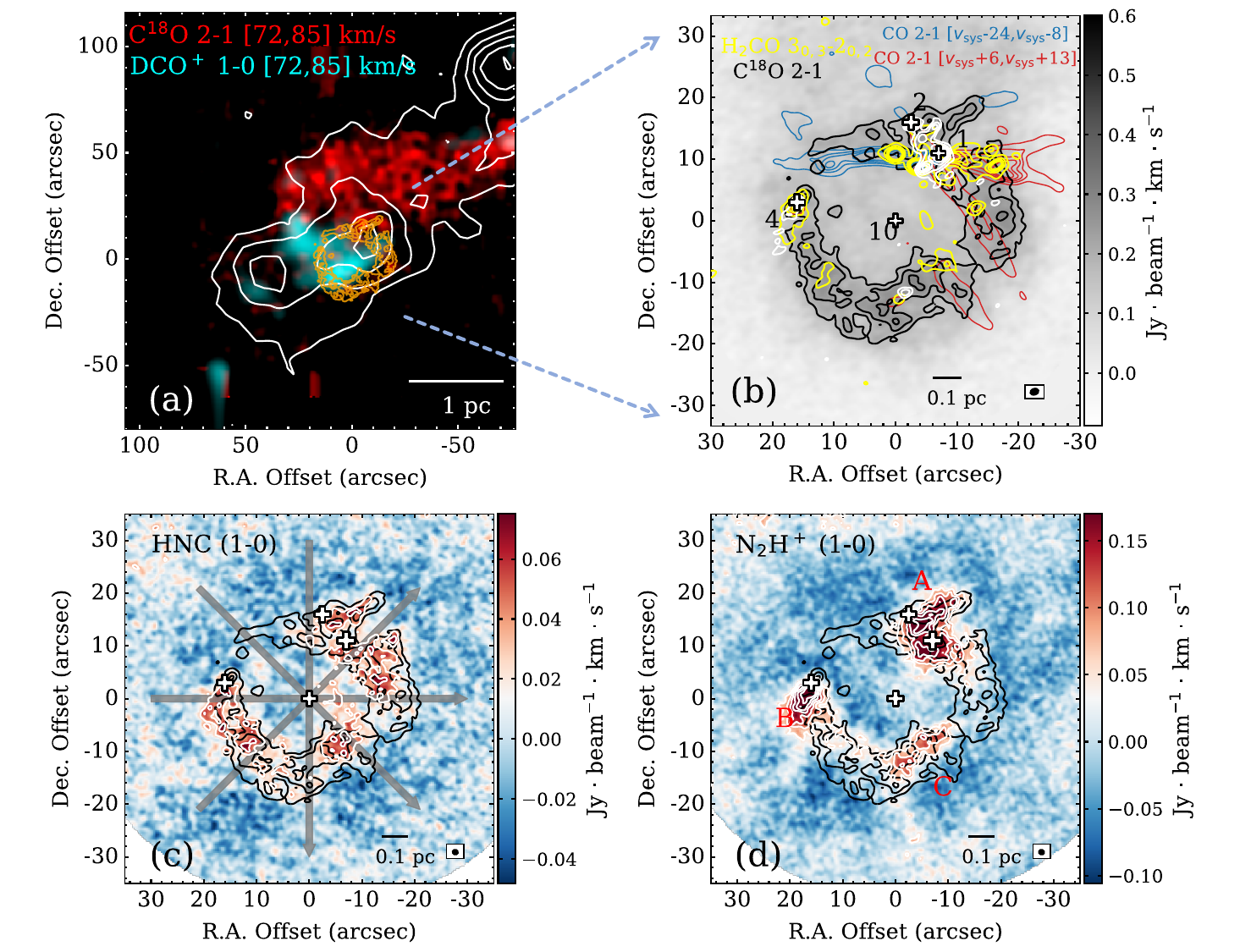}
    \caption{
    (a): Integrated intensity map of C$^{18}$O~(2–1) from ALMA 12m+7m+TP data with full UV coverage, integrated over the velocity range 79.9–81.9 km~s$^{-1}$, overlaid on large scale two-color maps of G34.74-0.12. 
    The red and cyan colors represent the integrated intensities of C$^{18}$O~(2–1) and DCO$^+$~(1–0), respectively, over the velocity range of 72-85~km~s$^{-1}$, obtained from IRAM-30m observations \citep{Feng2020ApJ...901..145F}.
    White contours represent the dust continuum emission from the combined Planck and APEX 870~$\mu$m map of the ATLASGAL survey, shown at contour levels of 4, 5, 6, 7, and 8~$\sigma$, where $\sigma = 0.18$Jy~beam$^{-1}$.
    The orange contours represents the C$^{18}$O~(2-1) gas ring-like structure with levels start from 10~$\sigma$ and increase in steps of 3~$\sigma$ (1~$\sigma$ =0.02 Jy beam$^{-1}$ km s$^{-1}$). The relative coordinate center (0,0) denotes the center of the ring. 
    (b): The greyscale map and black contours represent the integrated intensity of C$^{18}$O~(2-1) ring (integrated velocity range: 79.9 - 81.9 km~s$^{-1}$). The contour level are same as Fig. \ref{fig:C18O-ring} (a). 
    The blue and red contours represent the blue- and red-shifted components of outflows traced by CO~(2-1), respectively \citep{2025arXiv250714564L}. 
    The yellow contours represent the integrated intensity of H$_2$CO~(3$_{0,3}$-2$_{0,2}$) within the same integrated velocity range of C$^{18}$O~(2-1), showing a similar ring-like morphology. The contour levels of H$_2$CO start from 4~$\sigma$ and increase in steps of 4~$\sigma$ (1$\sigma$ =0.01 Jy beam$^{-1}$ km s$^{-1}$). The white contours represents the 1.3~mm dust continuum from ASHES observations. Black crosses mark the positions of core 1, 2, 4, and 10, as labeled by \cite{Morii2023ApJ...950..148M}. 
   (c):  C$^{18}$O~(2–1) ring overlaid on the integrated intensity map of HNC~(1-0). The black contours represents the C$^{18}$O~(2-1) gas ring-like structure, same as Fig. \ref{fig:C18O-ring} (a).
   The white contours represents the HNC~(1-0) emission with levels start from 3~$\sigma$ and increase in steps of 2~$\sigma$ (1~$\sigma$ = 0.012 Jy beam$^{-1}$ km s$^{-1}$), integrated over  the same integrated velocity range of C$^{18}$O~(2-1).
   The grey arrows indicate the paths along four different directions, corresponding to PA of 0$^{\circ}$, 45$^{\circ}$, 90$^{\circ}$, and 135$^{\circ}$. 
   (d): C$^{18}$O~(2–1) ring overlaid on the integrated intensity map of N$_2$H$^+$~(1-0), derived by stacking its hyperfine components.
   The black contours represents the C$^{18}$O~(2-1) gas ring-like structure, same as panel (a). The white contours represents the N$_2$H$^+$~(1-0) emission with levels start from 3~$\sigma$ and increase in steps of 2~$\sigma$ (1~$\sigma$ = 0.019 Jy beam$^{-1}$ km s$^{-1}$), integrated over  the same integrated velocity range of C$^{18}$O~(2-1).
   A, B, and C mark the positions where HNC~(1-0), N$_2$H$^+$~(1-0), and C$^{18}$O~(2–1) lines all show emission S/N $>$ 4. 
   The synthesized beam for each line is shown in the bottom right corner of each panel. 
   }
    \label{fig:C18O-ring}
\end{figure*}

\subsection{A Ring-like structure}
\label{subsection:ring}
At the southeastern tail of the filamentary IRDC G34.74–0.12, a ring-like structure is revealed by C$^{18}$O~(2-1), HNC~(1-0), and N$_2$H$^+$~(1-0) emission lines with signal-to-noise ratio (S/N) $>$ 4.
This ring is located in a dense ($\sim$ 10$^5$ cm$^{-3}$) and cold ($\sim$ 15~K) region where 
C$^{18}$O exhibits a depletion factor of 5 at pc scale, as observed by IRAM-30m at an angular resolution of 11.8$^{\prime \prime}$ (see Fig.1 in  \citealt{Feng2020ApJ...901..145F}).

From the C$^{18}$O~(2–1) channel maps (see Appendix~\ref{Appendix: B}), the ring-like structure is detected within the velocity range of 79.9–81.9~km~s$^{-1}$ (spanning three channels), which is redshifted by 1–2~km~s$^{-1}$ relative to the systemic velocity of 79.0~km~s$^{-1}$. 
In the intensity map of C$^{18}$O~(2-1) integrated over the above velocity range (Fig. \ref{fig:C18O-ring} b), the ring exhibits a diameter of 0.68~pc and an eccentricity value of 0.08, when fitted with an elliptical profile, using emission above the 10$\sigma$ level to exclude gas contributions from the filamentary bulk motion. The center of the ring is spatially coincident with the protostellar core labeled as Core10 (3.4~$M_{\odot}$) \citep{Morii2023ApJ...950..148M},  which is associated with an extremely young ($\sim$ 10$^3$ year) protostellar outflow \citep{2025arXiv250714564L}. At the same velocity range, the HNC~(1–0) and N$_2$H$^+$(1–0) emission above $>$ 3~$\sigma$ spatially coincides with that of C$^{18}$O~(2–1), as shown in Fig. \ref{fig:C18O-ring} panels (c) and (d), respectively.
H$_2$CO~(3$_{0,3}$–2$_{0,2}$) also shows emission above 4~$\sigma$ in this region. 
However, instead of tracing the complete ring-like structure, its spatial distribution covers only approximately 40\% of the ring, with the strongest emission located in the northwestern corner (yellow contour), where several young protostellar outflows have been detected \citep{2025arXiv250714564L}.

\section{Analysis \& Discussion}
\label{Sec:Analysis_Discussion}

\subsection{Kinematic structure of the ring}
\label{subsect:Kinematic_properties}
\begin{figure*}
    \centering
    \includegraphics[width=0.98\linewidth]{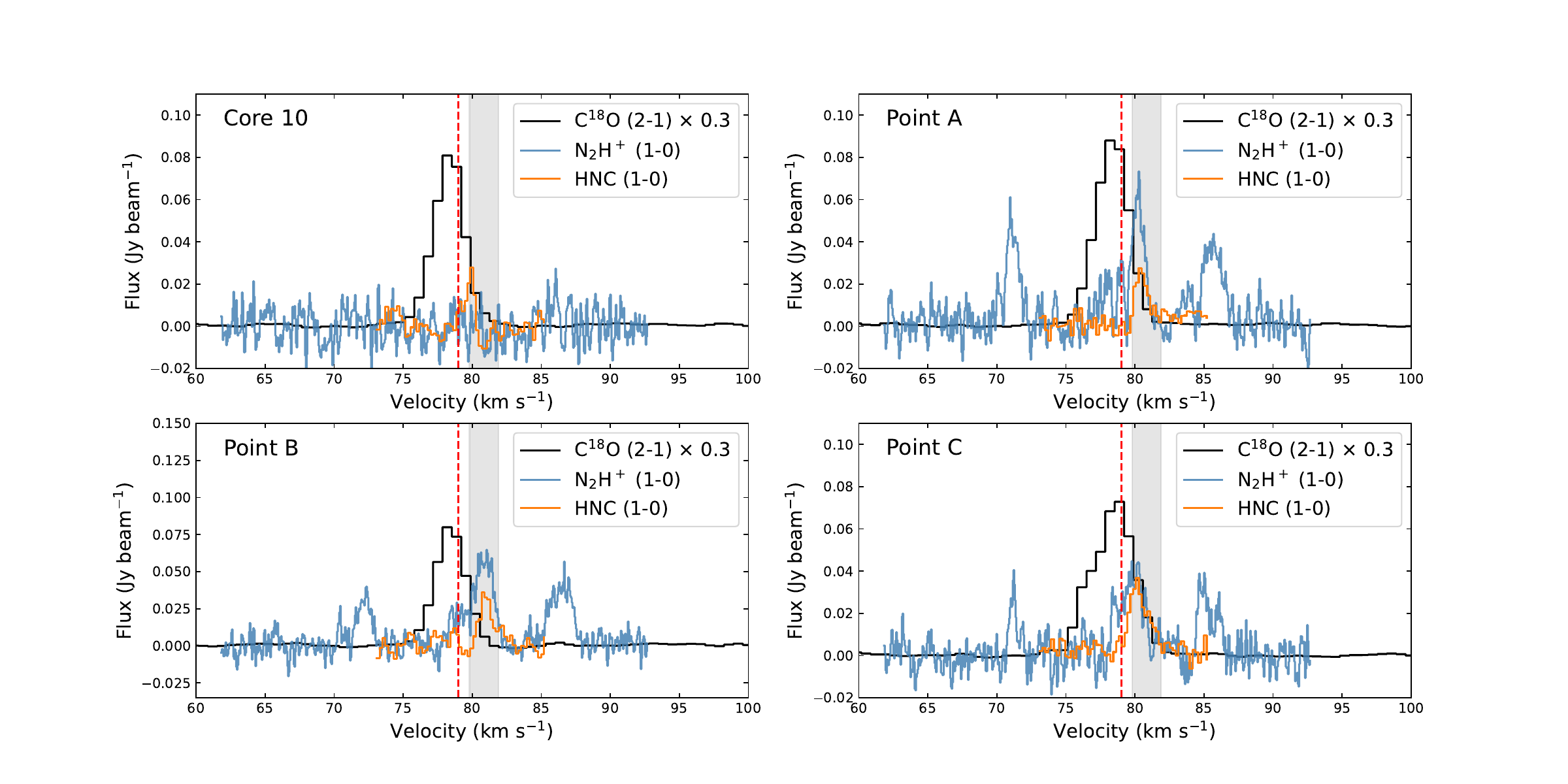}
    \caption{The line profiles of C$^{18}$O~(2–1), HNC~(1-0), and N$_2$H$^+$~(1-0), extracted from synthesized beam–averaged regions centered on core10, positions A, B, and C (as shown in Fig. \ref{fig:C18O-ring}). The vertical red dashed line indicates the systemic velocity, and the shaded region shows the integrated velocity range for the ring component.}
    \label{fig:Spectra_in_ring}
\end{figure*}

Within the ring-like structure, we extract the synthesized beam-averaged line profiles of C$^{18}$O~(2–1), N$_2$H$^+$(1–0), and HNC~(1–0) from three emission peaks located to the north (A), east (B), and south (C) of the ring (Fig.~\ref{fig:Spectra_in_ring}), where S/N $>$ 4 in Fig.~\ref{fig:C18O-ring}. Note that the HNC~(1–0) and N$_2$H$^+$ (1–0)~lines were obtained using interferometric data only. Therefore, the slight absorption features seen in their line profiles may result from missing short-spacing information. In contrast, the C$^{18}$O (2–1) line with complete uv coverage, shows broader line wings and a peak velocity that is offset by 1–2 km~s$^{-1}$ from those of HNC~(1-0) and N$_2$H$^+$~(1-0) (which peak at 80.3 km~s$^{-1}$).
For comparison, we also extract their beam-averaged line profiles toward the center of the ring (Core10, $V_{\rm sys,lsr}$ = 79 km~s$^{-1}$, \citealt{Morii2024ApJ...966..171M}). Although C$^{18}$O~(2-1) exhibits broader line wings with a Full Width at Half Maximum (FWHM) of 2~km~s$^{-1}$ toward this region, the S/N of the HNC~(1-0)and N$_2$H$^+$~(1-0) lines are $<$ 4 there.

The differences in peak velocity and line width arise from multiple factors. First, assuming a temperature of 20~K, the critical densities\footnote{The critical densities were derived under the assumption of optically thin conditions by solving the statistical equilibrium equations for a multi-level system, incorporating both downward collision rates ($\gamma_{ul}$) and excitation rates ($\gamma_{lu}$), using the \emph{myradex} package (https://github.com/fjdu/myRadex). The Einstein A coefficients ($A_{ij}$) and collisional rates ($C_{ij}$) are obtained from the Leiden atomic and molecular database (LAMDA; \citealt{Schoier2005AA...432..369S}).} of HNC~(1-0), N$_2$H$^+$~(1-0), and C$^{18}$O~(2-1) are $\sim$ 1.1 $\times$ 10$^5$ cm$^{-3}$, 6.0 $\times$ 10$^4$ cm$^{-3}$, and 4$\times$10$^3$ cm$^{-3}$, respectively, indicating that they trace gas with different densities. 
Second, the C$^{18}$O~(2-1) observations have lower angular resolution, and the N$_2$H$^+$~(1-0) transition exhibits hyperfine structures. As a result, both are less effective at resolving multiple kinematic components, such as ring and outflows, which are more clearly distinguished in the HNC (1–0) emission.
Therefore, in the following discussion, we primarily use the HNC~(1–0) line to investigate the kinematic properties of the ring.

\begin{figure*}
    \centering
    \includegraphics[width=0.9\linewidth]{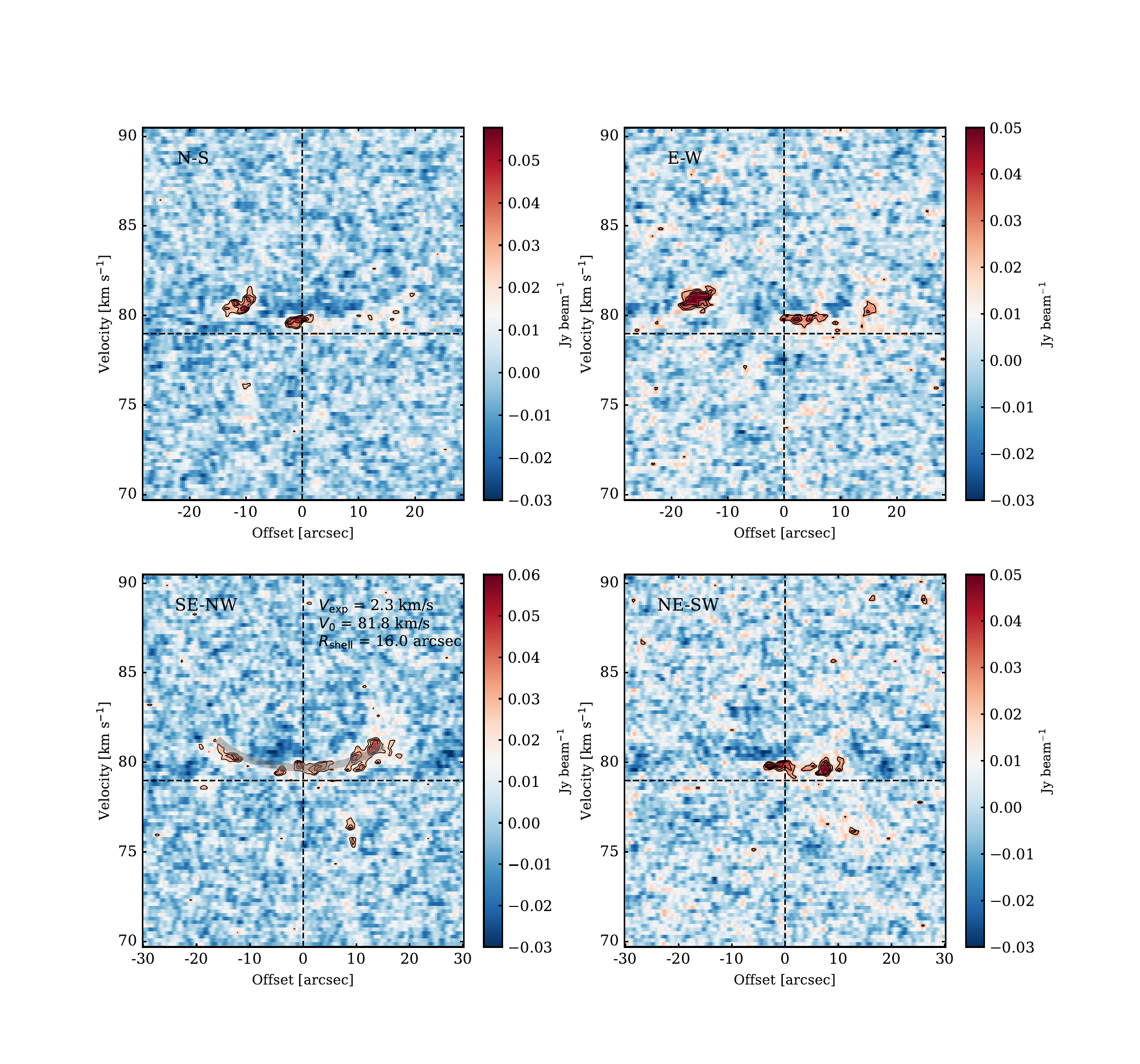}
    \caption{PV diagrams of HNC~(1-0) along four cuts with PA of 0$^{\circ}$, 45$^{\circ}$, 90$^{\circ}$, and 135$^{\circ}$ (labeled in Fig. \ref{fig:C18O-ring}) of the ring-like structure. Contour levels are set at 3, 4 and 5~$\sigma$, where $\sigma$=0.008~Jy~beam$^{-1}$. The vertical black dashed line marks the center of ring-like structure, while the horizontal black dashed line indicates the systemic velocity. The gray curve in the PV diagram represents the best-fit result from the spherical expanding shell model.
    }
    \label{fig:HNC-PV}
\end{figure*}

We extracted four position–velocity (PV) diagrams of HNC~(1–0) across the ring-like structure (Fig. \ref{fig:HNC-PV}). Each cut, with a width of beam size $\sim$ 1$\farcs$3, is centered on Core10 and oriented at PA of 0$^{\circ}$, 45$^{\circ}$, 90$^{\circ}$, and 135$^{\circ}$, respectively.
The PV diagrams in Fig.~\ref{fig:HNC-PV} reveal a ``U"-shaped feature along the SE–NW (PA: 45$^{\circ}$) cut, as indicated by the grey line, suggesting that the ring may exhibit signs of expansion. In contrast, no prominent ``U"- or ``V"-shaped structures are observed along the N–S (PA: 0$^{\circ}$), E–W (PA: 90$^{\circ}$), or NE–SW (PA: 135$^{\circ}$) directions. It remains unclear whether this is due to limited sensitivity or intrinsic asymmetries in the kinematics.
``U"- and ``V"-shaped structures in PV diagrams are commonly used as indicators of expansion signatures (e.g., \citealt{Arce2011ApJ...742..105A,Li2015ApJS..219...20L,Liu2024ApJ...964...93L}).

Assuming the gas is distributed on a thin spherical shell expanding at a constant velocity, the projected velocity structure is described by the following relation: 
\begin{equation}
    V(x) = V_0 \pm V_{\rm exp} \sqrt{1 - (\frac{x}{R_{\rm shell}})^2} ,
\end{equation}
where $V_0$ is the central velocity of the shell, $V_{\text{exp}}$ is the expansion velocity of the shell, and $R_{\rm shell}$ is the angular radius of the shell on the plane of the sky. We applied this simplified spherical expanding shell model to fit the observed PV diagram along SE-NW cut (PA = 45$^{\circ}$), chosen for its high signal-to-noise ratio. The best-fit parameters yield an expansion velocity ($V_{\text{exp}}$) of 2.3$\pm$2.6 km~s$^{-1}$, a systemic velocity ($V_0$) of 81.8$\pm$2.5 km~s$^{-1}$, and a shell radius ($R_{\rm shell}$) of 16$\farcs$0$\pm$5$\farcs$3 (i.e., 0.39 pc), suggesting that the structure likely represents an expanding gas shell. Note that the fitted expansion velocity exhibits a relatively large uncertainty at the 1$\sigma$ level.
The absence of clear expansion signatures in the PV diagrams suggests that the internal structure undergoes a relatively disordered diffusion process.

To examine how the ring-like structure connects to the ambient cloud, we also present the PV diagrams of C$^{18}$O~(2–1) and N$_2$H$^+$~(1–0) across the ring in Appendix \ref{Appendix: E}, using the same cut as for HNC~(1–0). 
In the PV diagram of C$^{18}$O~(2–1), the velocity increases from 79 km s$^{-1}$ near offset 0 arcsec to about 81 km s$^{-1}$ at offsets of $\pm$20 arcsec. 
The ring structure may gradually merge into surrounding extended, lower-density envelope.
However, the observed velocity difference is not prominent given the velocity resolution of 1.33~km s$^{-1}$. 
In contrast, N$_2$H$^+$~(1–0), which offers higher velocity resolution but traces only dense gas, shows no clear or continuous kinematic transition in its PV diagram.

\subsection{Physical structure of the ring}
\label{subsec:Physical structure}
Assuming local thermodynamic equilibrium (LTE) and optically thin, the column density of the molecule is calculated using the following equation \citep{Mangum2015PASP..127..266M}:
\begin{align}
    N_{\rm mol} = & \frac{3h}{8 \pi^3 S_{ij}\mu^2 g_u} Q \frac{\rm{exp} \mathit{(\frac{E_u}{k_{\rm B} T_{\rm ex}})}}{\rm{exp}(\mathit{\frac{h \nu}{k_{\rm B} T_{\rm ex}}}) - 1} \notag \\
    & \times \frac{W}{B_\nu(T_{\rm ex}) - B_\nu(T_{\rm CMB})},
    \label{equ:Nmol}
\end{align}
where $k_{\rm B}$ is the Boltzmann constant, $h$ is the Planck constant, $Q$ is the partition function, $B_{\nu}$ is the Planck function, $\nu$ is the rest frequency of the transition, $\mu$ is the electric dipole moment, $S_{ij}$ is the line strength, $E_u$ is the energy of the upper state, $g_u$ is the statistical weight of the upper level, 
and $T_{\rm CMB}$ refers to the cosmic microwave background temperature (2.73~K). The excitation temperature ($T_{\rm ex}$) is set to 15~K, based on the temperature map derived from VLA NH$_3$ (1,1) and (2,2) observations at lower spatial resolution of $\sim$ 5$^{\prime \prime}$ (Lin et al., in prep.). 
The velocity-integrated intensity $W$ is calculated over the range 79.9 – 81.9~km~s$^{-1}$ for the molecule, where the ring is detected.
The column density of HNC~(1-0) in the ring is on the order of 10$^{12}$~cm$^{-2}$, and that of C$^{18}$O~(2-1) is $\sim$ 10$^{15}$~cm$^{-2}$.

The total gas mass associated with the ring is subsequently estimated using the following equation:
\begin{equation}
    M = d^2X_{\rm mol}^{-1}\overline{m}_{\rm H_2} \int_{\Omega} N_{\rm mol}(\Omega)d\Omega, 
\end{equation}
where the $d$ is the source distance, $\Omega$ is the total solid angle, and the mean molecular weight per H$_2$ molecule ($\overline{m}_{\rm H_2}$), is taken to be 2.72~$m_{\rm H_2}$.
The ring mass is estimated using the C$^{18}$O~(2–1) line, which is optically thin and includes total flux. We adopt an abundance ratio with respect to H$_2$ ($X_{\rm mol}$) of 2 $\times$ 10$^{-7}$, taking into account the depletion of the cloud \citep{Feng2020ApJ...901..145F}, resulting in a total mass of 109~$M_\odot$. We also integrated the C$^{18}$O (2–1) emission over 74–79 km s$^{-1}$ to estimate the ambient gas mass, which excludes the velocities associated with the ring. Adopting the same temperature assumed for the ring, we estimate an ambient gas mass of 693~$M_\odot$. The ring accounts for only about one-seventh of this mass, indicating that it is not a dominant component with respect to the overall ambient gas distribution.

Considering the uncertainties from the assumed excitation temperature (20\%) and absolute flux calibration (10\%), the total uncertainty in the column density estimation can reach up to $\sim$ 30\%. Taking into account the additional uncertainty from the distance estimate (10\%), the mass estimation can reach up to $\sim$ 50\%.
Taking into account the presence of the protostar and outflow, adopting a higher excitation temperature of 50~K could increase the estimated mass and column density by a factor of a few (2-3), leading to an increase in the mass contrast between the ring and the ambient gas from $\sim$1/7 to $\sim$1/3.
Note that ambient gas contributes $\sim$50\% of the ring mass (estimated from off-ring C$^{18}$O(2–1) emission within the integrated velocity window), and the adopted low-velocity cutoff likely causes $\sim$20\% underestimation of the ring mass, based on a comparison between the integrated velocity range and the full width at zero intensity (FWZI) linewidth of HNC (1–0).

\subsection{Large-scale environment and Young stellar object}
\label{subsec:environment}
We examine the large-scale environment to investigate its potential influence on this ring. A middle-aged supernova remnant (SNR), W44, with an estimated dynamical age of 2 $\times$ 10$^4$ year \citep{Smith1985MNRAS.217...99S,Wolszczan1991ApJ...372L..99W}, is located near G34.74$-$0.12 in projection, at a kinematic distance of 3.0~kpc \citep{Caswell1975A&A....45..239C}. 
Several radio recombination line (RRL) clumps were identified near this region by \citet{Liu2024A&A...689A..29L}.
We checked the VLA 6~GHz data from the GLOSTAR survey \citep{Brunthaler2021A&A...651A..85B} and found no detectable free-free emission, with an upper limit of $<$ 2$\sigma$, where $\sigma$ = 1.2$\times$10$^{-4}$ Jy~beam$^{-1}$, indicating the absence of ionized gas.
The 1.3~GHz continuum data from the SARAO MeerKAT Galactic Plane Survey (SMGPS) \citep{Goedhart2024MNRAS.531..649G} also show no detectable free-free or non-thermal emission in this region, with an upper limit of $<$ 1$\sigma$, where $\sigma$ = 20 $\mu$Jy~beam$^{-1}$.
Considering the complex Galactic environment, including the nearby spiral arms and bar structure, which introduces substantial uncertainties in the kinematic distance measurement, together with
the potentially strong dynamical impact of the supernova explosion (e.g., turbulence injected into the molecular cloud; \citealt{Kortgen2016MNRAS.459.3460K,Ganguly2024MNRAS.528.3630G}), 
we cannot completely rule out a possible association between the ring-like structure and SNR W44, even though W44 seems to have only a limited impact on it. 
Future observations targeting more molecular lines will be crucial to further test this scenario.

Furthermore, we examined whether this region hosts YSOs by comparing archival infrared images from Spitzer (3.6, 4.5, 5.8, 8.0 and 24~$\mu$m), Herschel (70~$\mu$m), and WISE (3.4, 4.6, 12, and 22~$\mu$m) with the YSO catalog provided by \cite{McBride2021AJ....162..282M}. No compact sources were found at the corresponding locations (Appendix \ref{Appendix: A}), suggesting that more mature YSOs are likely absent from this region. 

At millimeter wavelengths, the dust cores identified from 1.3~mm continuum, are predominantly distributed around the gas ring, as shown in Fig. \ref{fig:C18O-ring} (b).
In particular, at a scale of 5000~au, we found signatures of early star formation, as evidenced by at least four protostellar outflows with estimated dynamical timescales of $10^3$–$10^4$ years \citep{2025arXiv250714564L}, all spatially coincident with the ring structure (Fig. \ref{fig:C18O-ring} (b)). 
These outflows are associated with Core1 (9.7~$M_{\odot}$), Core2 (6.4~$M_{\odot}$), Core4 (4.1~$M_{\odot}$), and Core10 (3.4~$M_{\odot}$), as labeled by \citet{Morii2023ApJ...950..148M} and marked with black crosses in Fig. \ref{fig:C18O-ring}. 
Although the 3.0~mm continuum (with a 1$\sigma$ mass sensitivity of 0.9~$M_{\odot}$, using the method described in Appendix~\ref{Appendix: C}) is less sensitive than the 1.3~mm continuum (1$\sigma$ sensitivity of 0.1~$M_{\odot}$) for detecting deeply embedded dust cores, we are still able to identify several cores at the 3~mm emission peaks.
We estimate the mass of Core1 as 13.3~$M_{\odot}$ from 3.0~mm emission. The slight difference may be the result of the opacity assumption \citep{Roy2013ApJ...763...55R} and filtering factor of different wavelengths.

\subsection{CO Depletion}
\label{subsec:depletion}

Gaseous CO can be efficiently freeze out onto dust grain surfaces at temperature $\lesssim$ 15~K \citep[e.g.,][]
{Kramer1999A&A...342..257K,Bergin2002ApJ...570L.101B,Caselli2008A&A...492..703C,Wiles2016MNRAS.458.3429W,Sabatini2019MNRAS.490.4489S,Feng2020ApJ...901..145F}.
The depletion factor of gaseous CO ($f_{\rm D}$) is defined the ratio of the expected CO abundance with respect to H$_2$ to the observed value \citep[e.g.,][]{Caselli1999ApJ...523L.165C,Fontani2012MNRAS.423.2342F,Sabatini2019MNRAS.490.4489S}. 
Since the low-J $^{12}$C$^{16}$O lines are usually optically thick in dense IRDC environments, we use the C$^{18}$O (2–1) line, which can be assumed to be optically thin, to estimate the CO depletion factor (see details in  Appendix \ref{Appendix: D}).

Parsec-scale CO depletion towards G34.74-0.12 has been derived by \citet{Feng2020ApJ...901..145F}, reporting $f_{\rm D}$ of $\sim$ 5. The observed central CO depletion is likely caused by freeze-out onto dust grains, due to the low temperature (15~K) and high density (10$^5$ cm$^{-3}$) of this region. Although large-scale CO depletion may contribute to the observed ring-like structure,
we consider it unlikely to be the main cause, since the structure is also evident in dense gas tracers such as HNC (1–0) and N$_2$H$^+$ (1–0).
Although a clear anti-correlation between C$^{18}$O~(2–1) and DCO$^+$~(1–0) is evident in the IRAM-30m data when both are integrated over the same velocity range (Fig.~\ref{fig:C18O-ring} a), DCO$^+$ (1–0) cannot resolve the ring-like structure embedded within the clump, because the IRAM-30m observations have a limited spatial resolution of 36$^{\prime \prime}$ and a velocity resolution of 0.83 km s$^{-1}$.
Therefore, without higher angular resolution observations of this line, it is difficult to determine the relationship between DCO$^+$~(1–0) and ring-like structure in terms of both velocity and spatial structure.

To estimate the core-scale depletion, we integrated the C$^{18}$O (2–1) emission over the velocity range [75.1, 82.6] km s$^{-1}$, using the 1.3 mm continuum and C$^{18}$O (2–1) data from the combined 12m and 7m arrays.
According to the temperature map (10–18 K) derived from our VLA NH$_3$~(1,1) and (2,2) data, we find that all of these cores exhibit high depletion factors, with $f_\mathrm{D}$ ranging from 10 to 110, consistent with core-scale depletion observed in high-mass star-forming regions \citep[e.g.,][]{Zhang2009ApJ...696..268Z,Morii2021ApJ...923..147M,
Sabatini2022ApJ...936...80S}.
However, since the NH$_3$ (1,1) and (2,2) transitions are only sensitive to temperatures below $\sim$40 K  \citep{Walmsley1983A&A...122..164W}, we also derived $f_\mathrm{D}$ assuming a higher temperature of 50 K, which provide the depletion factor of 9-63 on the core scale.
The large $f_{\rm D}$ values associated with these cores may primarily caused by their increased average gas density relative to the clump average density.
Such elevated densities can promote dust grain coagulation and produce larger grains that absorb radiation less efficiently \citep{Galametz2019A&A...632A...5G}, thereby reducing grain heating and CO evaporation from their surfaces \citep{Iqbal2018A&A...615A..20I}.

Note that this estimation carries significant uncertainties, arising from (i) the uncertain gas temperature mentioned above, (ii) the assumed C$^{18}$O abundance with respect to H$_2$ (Equation \ref{equ:XE_C18O} in Appendix \ref{Appendix: D}), which is based on the canonical $^{16}$O/$^{18}$O ratio \citep{Wilson1994ARA&A..32..191W} and the $^{12}$C$^{16}$O/H$_2$ ratio, assuming that CO abundance is primarily determined by the elemental C/H ratio \citep{Luck2011AJ....142..136L}, and (iii) missing flux in C$^{18}$O~(2–1) ($\sim$ 60\%), estimated from the IRAM-30m data, as well as in the dust continuum, which cannot be quantified from the available bolometric data.

\subsection{Possible origins of Ring-like Structure}

Ring-like structures have been reported in various star-forming regions, with most of them identified as infrared or radio bubbles driven by expanding H\,\textsc{ii} regions or stellar winds from massive stars \citep{Churchwell2006ApJ...649..759C,Li2019MNRAS.487.1517L,Watkins2023ApJ...944L..24W,Shen2025A&A...693A..21S}. 
The morphology, diameter, and mass of the ring-like structure found in G34.74–0.12 are broadly comparable to those of molecular shells observed around compact H\,\textsc{ii} regions \citep[e.g.,][]{Zhang2024MNRAS.535.1364Z} or YSOs \citep[e.g.,][]{Feddersen2018ApJ...862..121F}. Note that the observed ring also shares morphological similarities with the ``magnetic ring”, for which possible observational evidence has been reported only by \cite{Tokuda2023ApJ...956L..16T}. 

As mentioned in Section~\ref{subsec:environment}, although the northern, western, and southern parts of the ring are spatially associated with several identified protostellar outflows, no detectable free-free emission or H\,\textsc{ii} regions are found in this 70~$\mu$m-dark region. Moreover, the CO depletion factor in this region reaches up to 10 on clump (pc) scales.
This suggests that the ring may have a different formation mechanism compared to more evolved regions, potentially linked to early internal dynamics within the IRDC.
Four possible formation mechanisms may have led to the observed ring-like structure.

\textit{1. Relic of wind-blown bubble from YSO.}
Note that the ring-like structure found in our observation is morphologically similar to several expanding spherical shells driven by stellar winds around stars in both low-mass and high-mass star-forming regions (e.g., Taurus, Perseus, Orion A molecular cloud, and the central molecular zone (CMZ), \citealt{Arce2011ApJ...742..105A,Li2015ApJS..219...20L,Feddersen2018ApJ...862..121F,Henshaw2022MNRAS.509.4758H, Duan2023ApJ...943..182D}). These wind-blown bubbles have radii ranging from 0.05~pc to 3~pc, and the radius of our ring-like structure is consistent with this range. The detection of wind-blown bubbles in young star-forming regions remains relatively rare due to their features are less distinct compared to outflows.
In particular, only two previously reported bubbles coexisting with outflows \citep{Feddersen2018ApJ...862..121F,Duan2023ApJ...943..182D}. 
As mentioned in Section \ref{subsect:Kinematic_properties}, the ring observed in G34.74-0.12 exhibits only relatively weak signatures of expansion, but it may represent a relic of a young wind-blown bubble. Assuming that the structure is a bubble formed by stellar or disk winds originating from a deeply embedded YSO, the estimated bubble mass of 109~$M_{\odot}$ is comparable to the values reported for bubbles in Taurus \citep{Li2015ApJS..219...20L}. 
Based on the relation $t_{\rm dyn} = R_{\rm shell}/V_{\rm exp}$, we derive a dynamical timescale of 1.6 $\times$ 10$^5$ years for the bubble. This timescale is comparable to the free-fall timescale ($4 \times 10^5$~yr) of the clump and greater than the collapse timescale of the protostellar cores (2.6$\times$10$^4$ to 9.0$\times$10$^4$ yr) in this region. 
While the driving source of the bubble remains uncertain, both the high-mass Core1 and the centrally located Core10 are plausible candidates. 
This process will compress the surrounding gas and may lead to star formation around the ring.

\textit{2. Hollow structure result from the cloud interaction.}
Considering that this ring-like structure component has a narrow linewidth of 2~km~s$^{-1}$ and the entire ring is red-shifted relative to the systemic velocity (section \ref{subsection:ring}),
it resembles an outward-moving ring along the line of sight, likely resulting from non-isotropic expansion, rather than a uniformly diffusive spherical shell. 
No kinematic features indicative of interaction are detected in the C$^{18}$O~(2-1) PV diagram (Appendix~\ref{Appendix: E}), possibly due to the limited velocity resolution (1.33 km$^{-1}$).
Large-scale maps of the $^{12}$CO~(1-0), $^{13}$CO~(1-0), and C$^{18}$O~(1-0) from the MWISP survey \citep{Su2019ApJS..240....9S} also show no clear signs of interaction.
Nevertheless, the IRAM-30m line maps reveal a clear velocity gradient, from 77.5 km~s$^{-1}$ at a declination offset of +50 arcsec to 79 km~s$^{-1}$ at -50 arcsec (Fig. \ref{fig:C18O-ring} a).
Therefore, the red-shifted ring-like structure may still represent the relic resulting from interactions between clouds, such as filamentary structure colliding and acrossing the dense clouds.
The gas ring may have formed as a result of a molecular cloud passing through a pre-existing dense region, triggering gravitational instabilities and shock compression.
Simulations by \citet{Takahira2014ApJ...792...63T} show that the collision of a small cloud with a larger one can produce a cavity or ring-like structure in the larger cloud. 
They suggested that this process may result in the accumulation of dense gas along the interaction front, which in turn may trigger localized star formation. 
Observational evidence of such structures has been reported in several massive star-forming regions (e.g., \citealt{Torii2015ApJ...806....7T,Fukui2018ApJ...859..166F,Maity2025AJ....169..325M}).
The detection of multiple protostellar cores at 1.3~mm (Fig. \ref{fig:C18O-ring} b) and 3.0~mm, along with their associated outflows coinciding with the ring-like structure, suggesting that the ring observed in G34.74-0.12 may have formed in such a scenario.

\textit{3. Gas ring produced by temperature gradient.}
Note that, at a spatial resolution of 0.3-0.7 pc, we did not observe the same ring in the C$^{18}$O~(2-1) map within the same velocity range, neither in the ALMA TP data nor in the IRAM-30m data. Considering that this 70~$\mu$m dark clump has several extremely young embedded protostars and exhibits significant CO depletion at pc scale, with enhanced DCO$^+$~(1-0) (Fig. \ref{fig:C18O-ring} a), we propose that this ring could be an indicator of the temperature transition zone, i.e., outward from the ring (from the spatial extension of 0.34 pc),  CO adheres to the cold envelope ($\sim$ 15\,K, \citealt{Feng2020ApJ...901..145F}) where DCO$^+$ remains in the gas phase, while within the ring, UV radiation from young stellar objects has destroyed the molecular gas. 
However, our current NH$_3$ data lack sufficient spatial resolution and temperature sensitivity to test this temperature gradient. Future observations of C$^+$ and [C\textsc{I}] will be required to test it.

\textit{4. Line-of-sight coincidence of multiple outflows or dense clouds.}
Considering that there are multiple outflows present near this ring (Fig. \ref{fig:C18O-ring} (b)), we cannot rule out the possibility that the observed ring is a line-of-sight superposition of multiple outflows or dense cloud, similar to the complex ring-like structures discussed in \citet{Nissen2012A&A...540A.119N,
Henshaw2016MNRAS.457.2675H}. Moreover, molecular outflows traced by HNC and N$_2$H$^+$ are rare, with only one reported case to date \citep{Morii2025ApJ...979..233M}.

\section{Conclusion} 
\label{sec:summary}
We discover a dense molecular gas ring in the 70~$\mu$m dark region G34.74–0.12, revealed by S/N $>$ 4 emission of C$^{18}$O (2-1), HNC (1-0), and N$_2$H$^+$ (1-0) lines, based on ALMA observations at a linear resolution of $\sim$ 5000~au. 

With a diameter of 0.68~pc and a total mass of 109~$M_\odot$, this redshifted ring ($V_{\rm sys,lsr}$+0.9 to $V_{\rm sys,lsr}$+2.9~km~s$^{-1}$) is spatially coincident with a large-scale CO depletion region,
the 1.3~mm and 3.0~mm dust continuum emission, as well as several protostellar outflows. The column densities of C$^{18}$O~(2-1) and HNC~(1-0) toward the ring are estimated to be 10$^{15}$ cm$^{-2}$ and 10$^{12}$ cm$^{-2}$, respectively.
Neither detectable free-free emission nor any associated H\textsc{ii} region is found within this 70~$\mu$m-dark and CO depleted region. The PV diagram along the SE–NW cut reveals a ``U"-shaped structure, suggesting a possible expansion of the ring. Assuming an expanding shell model, we estimate an expansion velocity of 2.3~km~s$^{-1}$ and a dynamical timescale of approximately 10$^5$ years. All observational properties suggest that this region is in an early evolutionary stage, yet star formation activity has already begun.

Four possible origins may result in this ring-like structure, including a line-of-sight coincidence of multiple outflows or dense clouds, a gas ring produced by temperature gradient, a remnant of a wind-blown bubble, or a hollow structure result from cloud–cloud interactions.  
The latter two scenarios could potentially trigger subsequent star formation around the ring. 
Further observations will be essential to distinguish among the above scenarios and reveal the star formation activity associated with the ring.
\\



\begin{acknowledgments}
    We thank Paul Ho for the constructive comments and suggestions, which have greatly improved this manuscript.
    This letter makes use of the ALMA data ADS/JAO.ALMA\#2017.1.00716.S (PI: P.~Sanhueza) and \#(2022.1.01203.S (PI: Mardones Diego). ALMA is a partnership of ESO (representing its member states), NSF (USA) and NINS (Japan), together with NRC (Canada), MOST and ASIAA (Taiwan), and KASI (Republic of Korea), in cooperation with the Republic of Chile. The Joint ALMA Observatory is operated by ESO, AUI/NRAO and NAOJ.
    S.L. and S.F. acknowledge support from the National Key R\&D program of China grant (2025YFE0108200), National Science Foundation of China (12373023,  1213308), the starting grant at Xiamen University, and the presidential excellence fund at Xiamen University. KW acknowledges support from the National Science Foundation of China (12033005), the Tianchi Talent Program of Xinjiang Uygur Autonomous Region, the China-Chile Joint Research Fund (CCJRF; No. 2211). PS was partially supported by a Grant-in-Aid for Scientific Research (KAKENHI Number JP22H01271 and JP23H01221) of JSPS.
    GS acknowledges the project PRIN-MUR 2020 MUR BEYOND-2p (``Astrochemistry beyond the second period elements'', Prot. 2020AFB3FX), the PRIN MUR 2022 FOSSILS (``Chemical origins: linking the fossil composition of the Solar System with the chemistry of protoplanetary disks'', Prot. 2022JC2Y93), the project ASI-Astrobiologia 2023 MIGLIORA (``Modeling Chemical Complexity'', F83C23000800005), the INAF-GO 2023 fundings PROTOSKA (``Exploiting ALMA data to study planet forming disks: preparing the advent of SKA'', C13C23000770005) and the INAF-Minigrant 2023 TRIESTE (“TRacing the chemIcal hEritage of our originS: from proTostars to planEts”; PI: G. Sabatini). HL is supported by the National Key R\&D Program of China No. 2023YFB3002502, the National Natural Science Foundation of China under No. 12373006 and 12533004, and the China Manned Space Program with grant No. CMS-CSST-2025-A10.
\end{acknowledgments}

\begin{contribution}
S.L. carried out the data reduction and analysis under the supervision of S.F.. S.F. conceived the project, coordinated the research activities, and served as the corresponding author. S.P. led the ASHES project. K.W. led the VELLA project. F.X. was responsible for the data reduction of VELLA data. 
All authors discussed the results and contributed to the final manuscript.


\end{contribution}

%
\facilities{The Atacama Large Millimeter/submillimeter Array (ALMA).}

\software{CASA \citep{McMullin2007ASPC..376..127M}, APLpy \citep{aplpy2012,aplpy2019}, \href{http://www.astropy.org}{Astropy} (\citealt{Astropy2013A&A...558A..33A}), Matplotlib \citep{Hunter2007CSE.....9...90H}.
          }



\newpage
\appendix
\setcounter{section}{0}
\renewcommand{\thetable}{A\arabic{section}}
\setcounter{table}{0}
\renewcommand{\thetable}{A\arabic{table}}
\setcounter{figure}{0}
\renewcommand{\thefigure}{A\arabic{figure}}

\section{Channel map}
\label{Appendix: B}
\begin{figure*}[htb!]
\centering
\includegraphics[width=\linewidth]{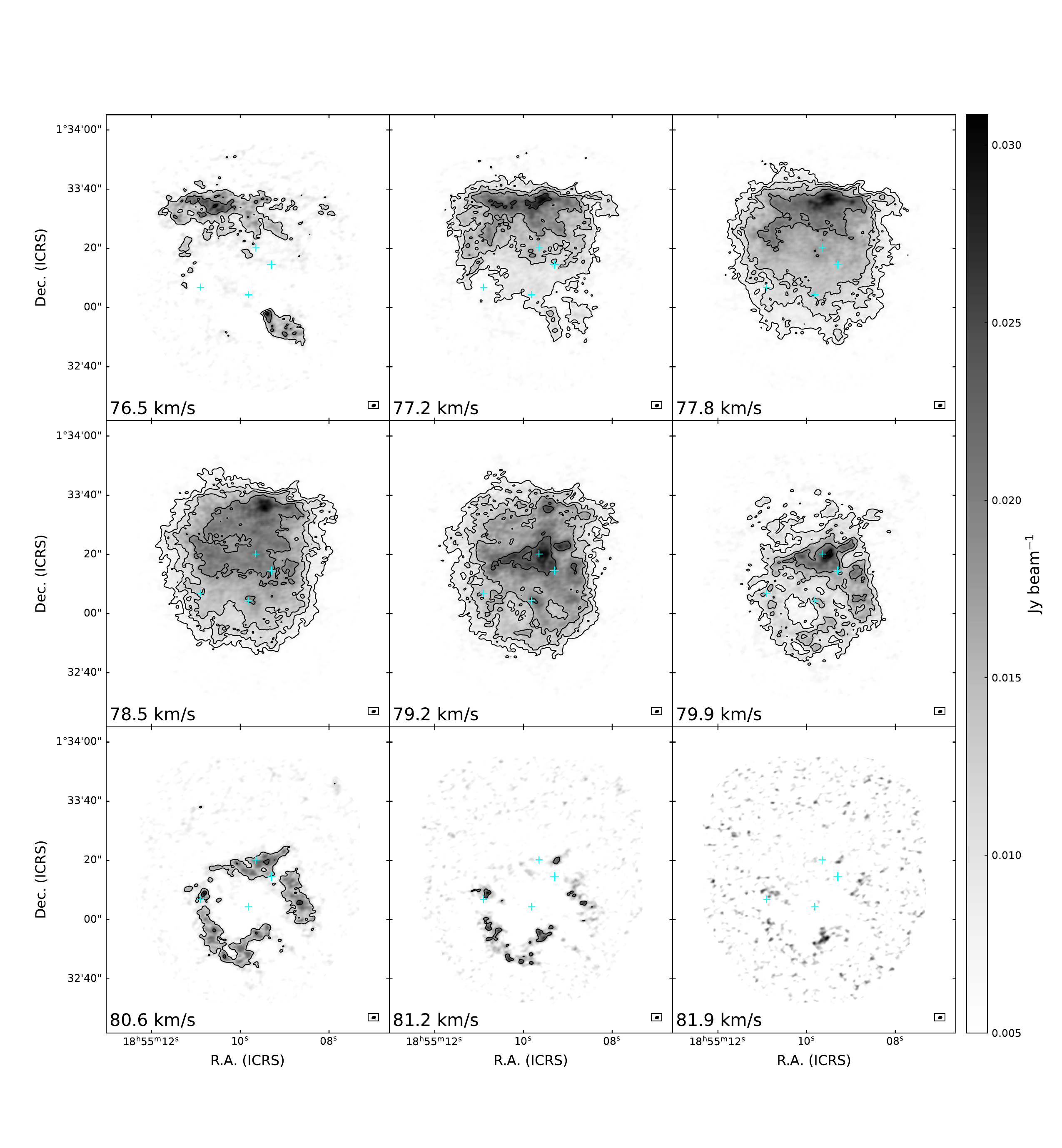}
\caption{C$^{18}$O~(2-1) channel map. The contour levels start from 5~$\sigma$ and increase in steps of 6~$\sigma$ (1~$\sigma$ =0.007 Jy beam$^{-1}$). The cyan crosses indicate the position of core 1, 2, 4, and 10, as labeled in Fig. \ref{fig:C18O-ring}. The synthesized beam is given in the bottom right.}
\label{fig:C18O_channel_map}
\end{figure*}

\begin{figure*}[htb!]
    \centering
    \includegraphics[width=\linewidth]{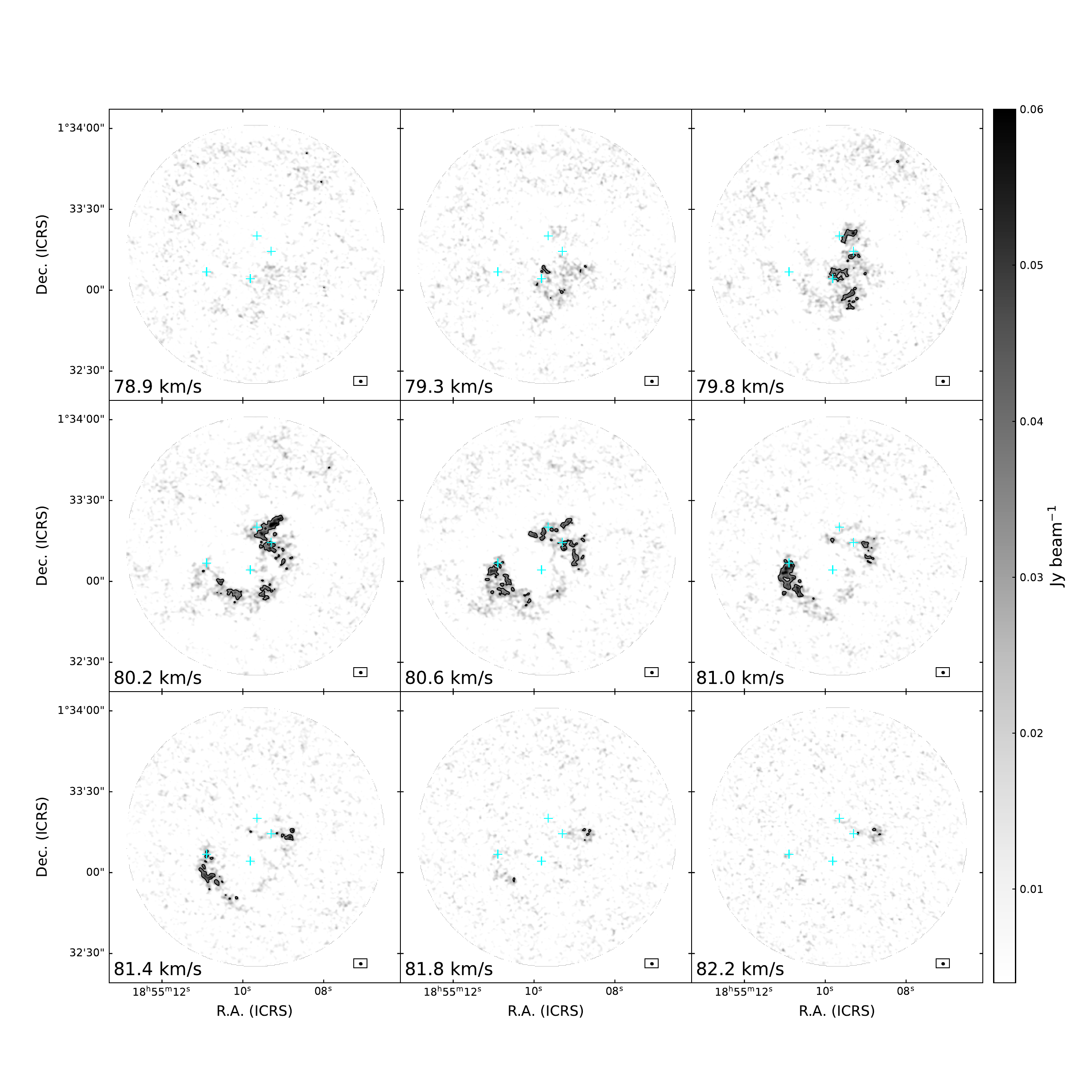}
    \caption{HNC~(1-0) channel map. The observational velocity resolution is 0.21~kms$^{-1}$; here, we smooth the data to 0.42~kms$^{-1}$. The contour levels start from 4~$\sigma$ and increase in steps of 10~$\sigma$ (1~$\sigma$ =0.008 Jy beam$^{-1}$). The cyan crosses indicate the position of core 1, 2, 4, and 10, as labeled in Fig. \ref{fig:C18O-ring}. 
    The synthesized beam is given in the bottom right.}
    \label{fig:HNC_channel_map}
\end{figure*}

Figure \ref{fig:C18O_channel_map} and \ref{fig:HNC_channel_map} present the channel maps of C$^{18}$O~(2–1) and HNC~(1–0), respectively.

\section{Continuum images}
\label{Appendix: A}
\begin{figure*}
    \centering
    \includegraphics[width=\linewidth]{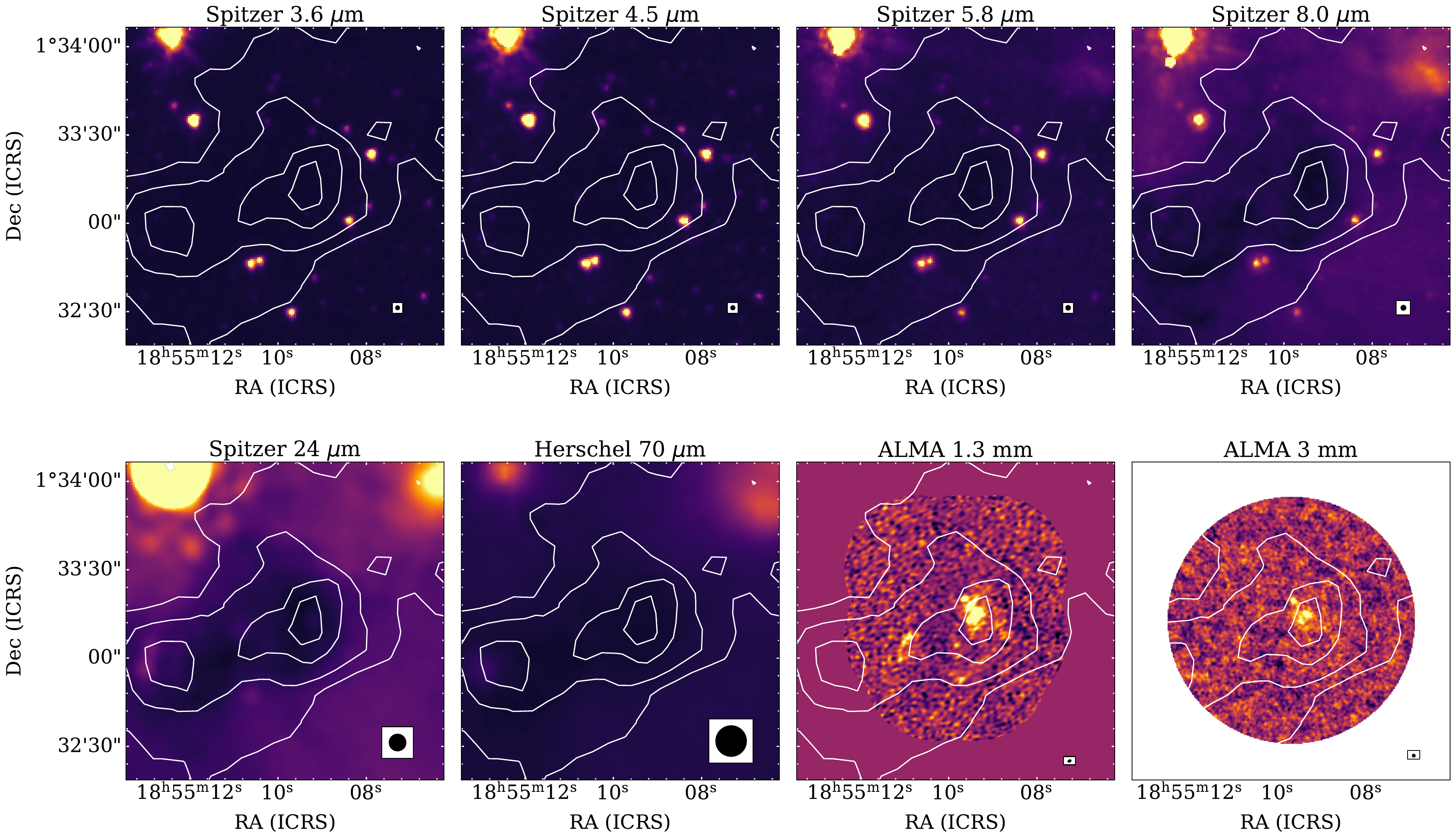}
    \caption{Multi-wavelength continuum images in the range of 3.6~$\mu$m to 3~mm. Top row: Spitzer 3.6~$\mu$m, 4.5~$\mu$m, 5.8~$\mu$m, and 8.0~$\mu$m images from GLIMPSE \citep{Churchwell2009PASP..121..213C}.
    Bottom row: Spitzer 24~$\mu$m image from MIPSGAL \citep{Carey2009PASP..121...76C}, Herschel 70~$\mu$m image, ALMA 1.3~mm image from ASHES \citep{Morii2023ApJ...950..148M}, and ALMA 3~mm image (this work). White contours indicate the dust continuum from the combined Planck/HFI and APEX/LABOCA map of the ATLASGAL survey, shown at levels of 4, 5, 6, 7 and 8~$\sigma$, where $\sigma$ = 0.18 Jy beam$^{-1}$. The angular resolution is shown at the bottom of each panel.}
    \label{fig:Mutilwave}
\end{figure*}

Figure \ref{fig:Mutilwave} presents multi-wavelength continuum images of G34.74$-$0.12.

\section{Core mass estimation}
\label{Appendix: C}
The core mass is estimated from the 3~mm dust continuum emission, assuming optically thin thermal dust emission and local thermodynamic equilibrium (LTE). The mass is calculated using the following equation:
\begin{equation}
    M = \frac{D^2 S_{\nu}}{\kappa_{\nu} B_{\nu}(T_{\rm dust})}  .
\end{equation}
where $D$ is the distance to the source, $S_{\nu}$ is the integrated flux density, $\kappa_{\nu}$ is the dust opacity per unit gas mass, and $B_{\nu}(T_{\rm dust})$ is the Planck function at dust temperature $T_{\rm dust}$. We assume a dust temperature of 15~K, a gas-to-dust mass ratio of 100,
and adopt a dust opacity of $\kappa_{3\,{\rm mm}}$ = 0.18~${\rm cm}^2\,{\rm g}^{-1}$ at 3~mm \citep{Ossenkopf1994A&A...291..943O}. 

\section{PV diagrams of C$^{18}$O~(2-1) and N$_2$H$^+$~(1-0)}
\label{Appendix: E}
\begin{figure*}
    \centering
    \includegraphics[width=\linewidth]{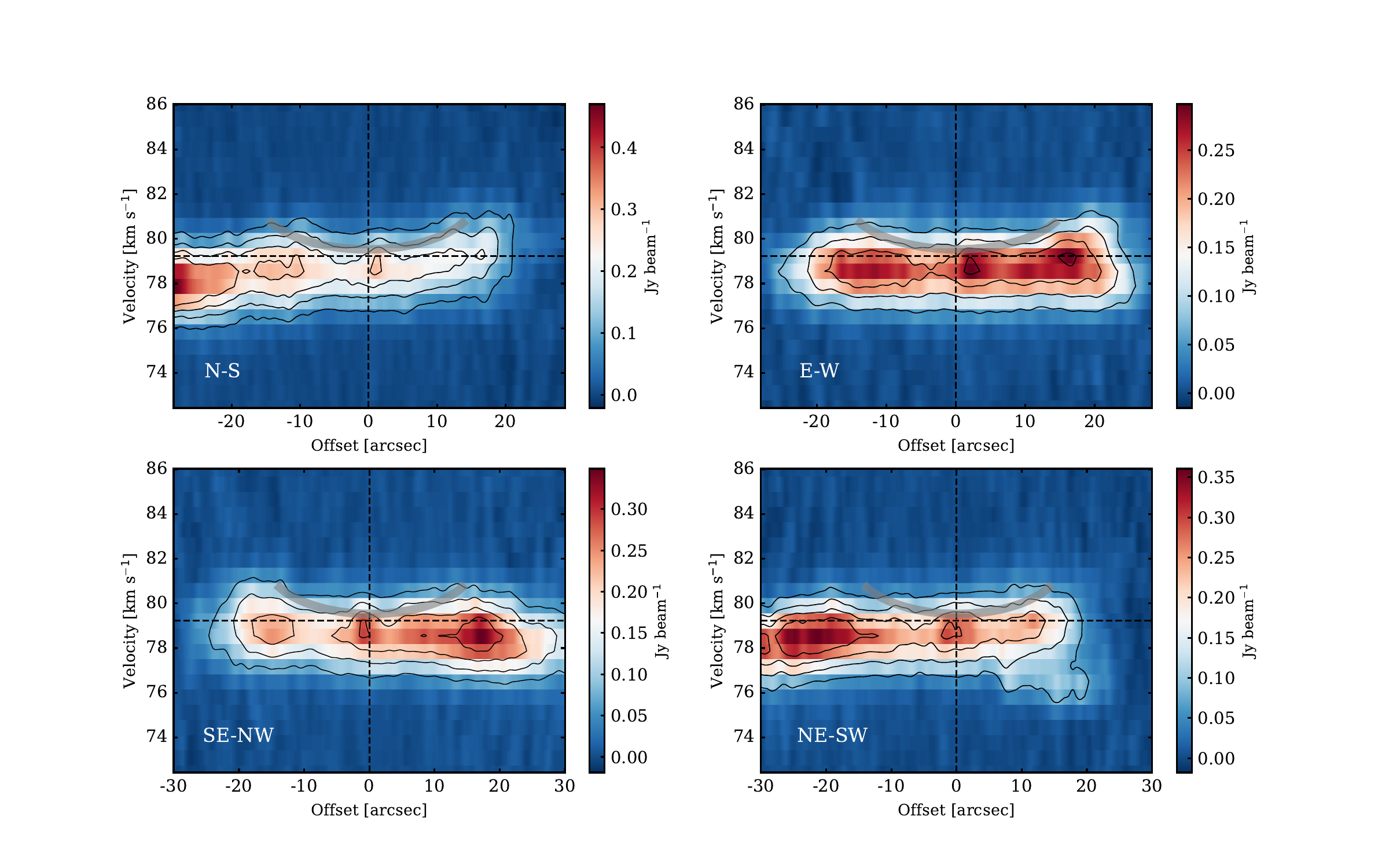}
    \caption{PV diagrams of C$^{18}$O~(2-1) along four cuts with PA of 0$^{\circ}$, 45$^{\circ}$, 90$^{\circ}$, and 135$^{\circ}$ (labeled in Fig. \ref{fig:C18O-ring}) of the ring-like structure. The gray lines indicate the fitting results from HNC~(1–0) (Fig. \ref{fig:HNC-PV}). Contour levels are set at 10, 20, 30 and 40~$\sigma$, where $\sigma$=0.007~Jy~beam$^{-1}$.}
    \label{fig:PV-C18O}
\end{figure*}

\begin{figure*}
    \centering
    \includegraphics[width=\linewidth]{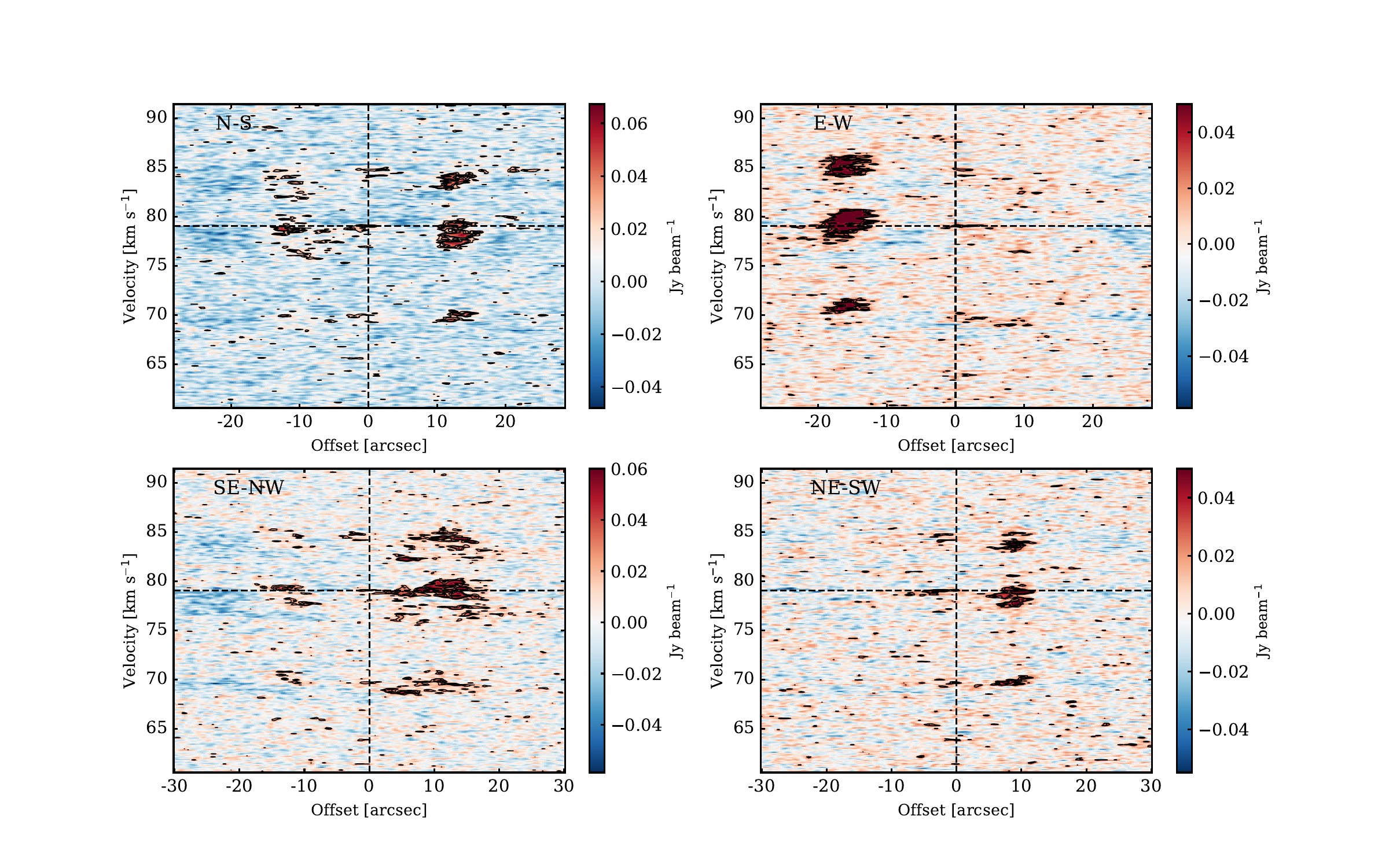}
    \caption{PV diagrams of N$_2$H$^+$~(1-0) along four cuts with PA of 0$^{\circ}$, 45$^{\circ}$, 90$^{\circ}$, and 135$^{\circ}$ (labeled in Fig. \ref{fig:C18O-ring}) of the ring-like structure. Contour levels are set at 3, 4, and 5~$\sigma$, where $\sigma$=0.008~Jy~beam$^{-1}$.}
    \label{fig:PV-N2Hp}
\end{figure*}

Figure \ref{fig:PV-C18O} and \ref{fig:PV-N2Hp} present the PV diagrams of C$^{18}$O~(2-1) and N$_2$H$^+$~(1-0), respectively.

\section{Depletion factor estimation}
\label{Appendix: D}
The expected C$^{18}$O/H$_2$ abundance ($X^{\rm E}_{\rm C^{18}O}$) is calculated as:
\begin{equation}
    X^{\rm E}_{\rm C^{18}O} = \frac{9.5 \times 10^{-5} \times 10^{\alpha(\frac{R_{\rm GC}}{\rm 1~kpc}-\frac{R_{\rm GC,\odot}}{\rm 1~kpc})}}{\rm [^{16}O]/[^{18}O]},
    \label{equ:XE_C18O}
\end{equation}
where $R_{\rm GC}$ is Galactocentric distance, adopted as 5.2~kpc for G34.74-0.12 \citep{Wenger2018ApJ...856...52W}, and $R_{\rm GC,\odot}$ is the solar Galactocentric distance, taken to be 8.34 kpc \citep{Reid2014ApJ...783..130R}. The oxygen isotopic ratio is derived using the relation $ \rm [^{16}O]/[^{18}O]$ = 58.8$R_{\rm GC}$ + 37.1 \citep{Wilson1994ARA&A..32..191W}, and $\alpha$ representing the radial carbon abundance gradient, taken to be -0.08 dex kpc$^{-1}$ \citep{Luck2011AJ....142..136L}. Using these values, we derived the $X^{\rm E}_{\rm C^{18}O}$ as 4.94$\times$10$^{-7}$.

The observed abundance of C$^{18}$O relative to H$_2$ ($X^{\rm O}_{\rm C^{18}O}$) is calculated as $N_{\rm C^{18}O}/N_{\rm H_2}$, where $N_{\rm C^{18}O}$ is derived using Equation~\ref{equ:Nmol},  based on the integrated intensity of the C$^{18}$O~(2-1) line over the velocity range [75.1,82.6] km~s$^{-1}$. The H$_2$ column density ($N_{\rm H_2}$) is derived from the following expression:
\begin{equation}
    N_{\rm H_2} = \frac{\gamma F_\nu}{B_\nu(T_{\rm dust})\kappa_\nu \mu_{\rm H_2} m_{\rm H_2} \Omega},
\end{equation}
where $F_\nu$ is flux density, $\gamma$ is gas-to-dust ratio, adopted as 78, estimated using the relation from \citet{Giannetti2017A&A...606L..12G}. $B_\nu(T_{\rm dust})$ is Planck function at dust temperature, $\kappa_\nu$ is the dust opacity, $\mu_{\rm H_2}$ is H$_2$ mean molecular weight, adopted as 2.8 \citep{Kauffmann2008A&A...487..993K}, $m_{\rm H_2}$ is the mass of an H$_2$ molecule, and $\Omega$ is the beam solid angle. The dust temperature is assumed to be equal to the gas temperature, which is derived from NH$_3$ observations and estimated to be 15~K (Lin et al. in prep; see Sect. \ref{subsec:Physical structure}). For 1.3~mm band, the adopted $\kappa_\nu$ is 0.9~cm$^2$~g$^{-1}$ \citep{Ossenkopf1994A&A...291..943O}.

The depletion factor of gaseous CO ($f_{\rm D}$) is then defined as the ratio of the expected abundance $X_{\rm C^{18}O}^E$ to the observed abundance $X_{\rm C^{18}O}^O$:
\begin{equation}
    f_D = \frac{X^{\rm E}_{\rm C^{18}O}}{X^{\rm O}_{\rm C^{18}O}}.
    \label{equ:depletion_factor}
\end{equation}


\bibliography{ref}{}
\bibliographystyle{aasjournalv7}



\end{document}